\documentstyle[12pt,twoside]{article}
 
\def\a{\alpha}
\def\b{\beta}

\def\d{\delta}
\def\e{\epsilon}                
\def\f{\phi}                    
\def\g{\gamma}
\def\h{\eta}

\def\k{\kappa}
\def\l{\lambda}
\def\m{\mu}
\def\n{\nu}
\def\o{\omega}
\def\p{\pi}                     
\def\r{\rho}                    
\def\s{\sigma}                  

\def\x{\xi}
\def\z{\zeta}
\def\D{\Delta}

\def\L{\Lambda}

\def\cf{{\cal F}}

\def\ci{{\cal I}}

\def\bo{{\raise.05ex\hbox{\large$\Box$}\:}}             
\def\cbo{{\,\raise-.15ex\Sc [\,}}                       
\def\pa{\partial}                                       
\def\su{\sum}                                           
\def\TH{{\raise.2ex\hbox{$\displaystyle \bigodot$}\mskip-4.7mu \llap H \;}}
\def\face{\hbox{\normalsize$\;\;\:{\raise.9ex\hbox{\oo n}\mskip-13mu \llap
        {${\buildrel{\hbox{\frtnrm ..}}\over\smile}$}}\:$}}     
\def\Face{{\raise.2ex\hbox{$\displaystyle \bigodot$}\mskip-2.2mu \llap {$\ddot
        \smile$}}}                                      
\def\Lhat{{\bf\rlap{\kern-.09em$\hat{\phantom L}$}L}}
\def\Lcheck{{\bf\rlap{\kern-.09em$\check{\phantom L}$}L}}
 
\def\sp#1{{}^{#1}}                              
\def\sbra#1{\left\langle #1\right|}             
\def\sket#1{\left| #1\right\rangle}             
\def\svev#1{\left\langle #1\right\rangle}       
\def\leftrightarrowfill{$\mathsurround=0pt \mathord\leftarrow \mkern-6mu
        \cleaders\hbox{$\mkern-2mu \mathord- \mkern-2mu$}\hfill
        \mkern-6mu \mathord\rightarrow$}
\def\dvec#1{\vbox{\ialign{##\crcr
        \leftrightarrowfill\crcr\noalign{\kern-1pt\nointerlineskip}
        $\hfil\displaystyle{#1}\hfil$\crcr}}}           
\def\ddt#1{{\buildrel {\hbox{\LARGE .\kern-2pt.}} \over {#1}}}
 
\def\frac#1#2{{\textstyle{#1\over\vphantom2\smash{\raise.20ex
        \hbox{$\scriptstyle{#2}$}}}}}                   
\def\ha{\frac12}                                        
\def\sfrac#1#2{{\vphantom1\smash{\lower.5ex\hbox{\small$#1$}}\over
        \vphantom1\smash{\raise.4ex\hbox{\small$#2$}}}} 
\def\bfrac#1#2{{\vphantom1\smash{\lower.5ex\hbox{$#1$}}\over
        \vphantom1\smash{\raise.3ex\hbox{$#2$}}}}       
\def\afrac#1#2{{\vphantom1\smash{\lower.5ex\hbox{$#1$}}\over#2}}    
 
\def\boxes#1{
        \newcount\num
        \num=1
        \newdimen\downsy
        \downsy=-1.64ex
        \mskip-7.8mu
        \bo
        \loop
        \ifnum\num<#1
        \llap{\raise\num\downsy\hbox{$\bo$}}
        \advance\num by1
        \repeat}
\def\boxup#1#2{\newcount\numup
        \numup=#1
        \advance\numup by-1
        \newdimen\upsy
        \upsy=.82ex
        \mskip7.8mu
        \raise\numup\upsy\hbox{$#2$}}
 
\newskip\humongous \humongous=0pt plus 1000pt minus 1000pt
\def\caja{\mathsurround=0pt}

\newif\ifdtup
\def\panorama{\global\dtuptrue \openup2\jot \caja
        \everycr{\noalign{\ifdtup \global\dtupfalse
        \vskip-\lineskiplimit \vskip\normallineskiplimit
        \else \penalty\interdisplaylinepenalty \fi}}}
\def\li#1{\panorama \tabskip=\humongous                         
        \halign to\displaywidth{\hfil$\displaystyle{##}$
        \tabskip=0pt&$\displaystyle{{}##}$\hfil
        \tabskip=\humongous&\llap{$##$}\tabskip=0pt
        \crcr#1\crcr}}

\def\PL{Phys. Lett. }
\def\PR{Phys. Rev. Lett. }
\def\PRD{Phys. Rev. D}
\def\CQG{Class. Quant. Grav.}

\def\ref#1{$\sp{#1]}$}

\parskip=\medskipamount                 
\def\baselinestretch{1.2}       
\thicklines                         
\def\title#1#2#3#4{
\begin{document}
        {\hbox to\hsize{#4 \hfill  #3}}\par
        \begin{center}\vskip.5in minus.1in {\Large\bf #1}\\[.5in minus.2in]{#2}
        \vskip1.4in minus1.2in {\bf ABSTRACT}\\[.1in]\end{center}
        \begin{quotation}\par}
\def\author#1#2{#1\\[.1in]{\it #2}\\[.1in]}

\def\AMIC{Aleksandar Mikovic\'c
\\[.1in]{\it Blackett Laboratory, Imperial College, Prince Consort Road, London
SW7 2BZ, UK}\\[.1in]}

\def\AMICIF{Aleksandar Mikovi\'c\,
\footnote{Work supported by MNTRS and Royal Society}
\\[.1in] {\it Blackett Laboratory, Imperial College, Prince Consort
Road, London SW7 2BZ, UK}\\[.1in]
and \\[.1 in]
{\it Institute of Physics, P.O. Box 57, 11001 Belgrade, Yugoslavia}
\footnote{Permanent address}\\ {\it E-mail:\, mikovic@castor.phy.bg.ac.yu}}

\def\AMSISSA{Aleksandar Mikovi\'c\,
\footnote{E-mail address: mikovic@castor.phy.bg.ac.yu}
\\[.1in] {\it SISSA-International School for Advanced Studies\\
Via Beirut 2-4, Trieste 34100, Italy}\\[.1in]
and \\[.1 in]
{\it Institute of Physics, P.O. Box 57, 11001 Belgrade, Yugoslavia}
\footnote{Permanent address}}

\def\AM{Aleksandar Mikovi\'c 
\footnote{E-mail address: mikovic@castor.phy.bg.ac.yu}
\\[.1in] {\it Institute of Physics, P.O.Box 57, Belgrade 11001, Yugoslavia}
\\[.1in]}

\def\AMsazda{Aleksandar Mikovi\'c 
\footnote{E-mail address: mikovic@castor.phy.bg.ac.yu}
and Branislav Sazdovi\'c \footnote{E-mail: sazdovic@castor.phy.bg.ac.yu}
\footnote{Work supported by MNTRS}
\\[.1in] {\it Institute of Physics, P.O.Box 57, Belgrade 11001, Yugoslavia}
\\[.1in]}

\def\AMVR{Aleksandar Mikovi\'c\,
\footnote{E-mail address: mikovic@castor.phy.bg.ac.yu}
\\[.1in] 
{\it Institute of Physics, P.O. Box 57, 11001 Belgrade, Yugoslavia}
\\[.2in]
Voja Radovanovi\'c \\[.1 in]
{\it Faculty of Physics, P.O. Box 550, 11001 Belgrade, Yugoslavia}}

\def\AMCVR{Aleksandar Mikovi\'c
\footnote{Permanent address: Institute of Physics, P.O. Box 57, 11001 
Belgrade, Yugoslavia}\footnote{E-mail: mikovic@fy.chalmers.se, 
mikovic@castor.phy.bg.ac.yu}
\\
{\it Institute of Theoretical Physics, Chalmers University of Technology,
S-412 96 Goteborg, Sweden}\\[.1in]
and
\\[.1in]
Voja Radovanovi\'c
\footnote{E-mail: rvoja@rudjer.ff.bg.ac.yu} \\
{\it Faculty of Physics, P.O. Box 550, 11001 Belgrade, Yugoslavia}}

\def\endtitle{\par\end{quotation}\vskip3.5in minus2.3in\newpage}
 
 
\def\endabstract{\par\end{quotation}
        \renewcommand{\baselinestretch}{1.2}\small\normalsize}
 
 
\def\xpar{\par}                                         

\def\letterhead{
        \centerline{\large\sf INSTITUTE OF PHYSICS}
        \centerline{\sf P.O.Box 57, 11001 Belgrade, Yugoslavia}
        \rightline{\scriptsize\sf Dr Aleksandar Mikovi\'c}
        \vskip-.07in
        \rightline{\scriptsize\sf Tel: 11 615 575}
        \vskip-.07in
        \rightline{\scriptsize\sf E-mail: MIKOVIC@CASTOR.PHY.BG.AC.YU}}

\def\sig#1{{\leftskip=3.75in\parindent=0in\goodbreak\bigskip{Sincerely yours,}
\nobreak\vskip .7in{#1}\par}}

\def\ssig#1{{\leftskip=3.75in\parindent=0in\goodbreak\bigskip{}
\nobreak\vskip .7in{#1}\par}}

 
\def\ree#1#2#3{
        \hfuzz=35pt\hsize=5.5in\textwidth=5.5in
        \begin{document}
        \ttraggedright
        \par
        \noindent Referee report on Manuscript \##1\\
        Title: #2\\
        Authors: #3}
 
 
\def\start#1{\pagestyle{myheadings}\begin{document}\thispagestyle{myheadings}
        \setcounter{page}{#1}}
 
 
\catcode`@=11
 
\def\ps@myheadings{\def\@oddhead{\hbox{}\footnotesize\bf\rightmark \hfil
        \thepage}\def\@oddfoot{}\def\@evenhead{\footnotesize\bf
        \thepage\hfil\leftmark\hbox{}}\def\@evenfoot{}
        \def\sectionmark##1{}\def\subsectionmark##1{}
        \topmargin=-.35in\headheight=.17in\headsep=.35in}
\def\ps@acidheadings{\def\@oddhead{\hbox{}\rightmark\hbox{}}
        \def\@oddfoot{\rm\hfil\thepage\hfil}
        \def\@evenhead{\hbox{}\leftmark\hbox{}}\let\@evenfoot\@oddfoot
        \def\sectionmark##1{}\def\subsectionmark##1{}
        \topmargin=-.35in\headheight=.17in\headsep=.35in}
 
\catcode`@=12
 
\def\sect#1{\bigskip\medskip\goodbreak\noindent{\large\bf{#1}}\par\nobreak
        \medskip\markright{#1}}
\def\chsc#1#2{\phantom m\vskip.5in\noindent{\LARGE\bf{#1}}\par\vskip.75in
        \noindent{\large\bf{#2}}\par\medskip\markboth{#1}{#2}}
\def\Chsc#1#2#3#4{\phantom m\vskip.5in\noindent\halign{\LARGE\bf##&
        \LARGE\bf##\hfil\cr{#1}&{#2}\cr\noalign{\vskip8pt}&{#3}\cr}\par\vskip
        .75in\noindent{\large\bf{#4}}\par\medskip\markboth{{#1}{#2}{#3}}{#4}}
\def\chap#1{\phantom m\vskip.5in\noindent{\LARGE\bf{#1}}\par\vskip.75in
        \markboth{#1}{#1}}
\def\refs{\bigskip\medskip\goodbreak\noindent{\large\bf{REFERENCES}}\par
        \nobreak\bigskip\markboth{REFERENCES}{REFERENCES}
        \frenchspacing \parskip=0pt \renewcommand{\baselinestretch}{1}\small}
\def\unrefs{\normalsize \nonfrenchspacing \parskip=medskipamount}
\def\Item{\par\hang\textindent}
\def\Itemitem{\par\indent \hangindent2\parindent \textindent}
\def\makelabel#1{\hfil #1}
\def\topic{\par\noindent \hangafter1 \hangindent20pt}
\def\Topic{\par\noindent \hangafter1 \hangindent60pt}

\include{psfig}
\title{Two-loop Back-reaction in 2D Dilaton Gravity} 
{\AMCVR}{Goteborg-ITP-96-7}{June 1996}

We calculate the two-loop quantum corrections, including the back-reaction
of the Hawking radiation, to the one-loop effective metric in a unitary
gauge quantization of the CGHS model of 2d dilaton gravity.
The corresponding evaporating black hole
solutions are analysed, and consistent semi-classical geometries 
appear in the weak-coupling region of the spacetime
when the width of the matter
pulse is larger then the short-distance cutoff. A consistent semi-classical
geometry also appears in the limit of a shock-wave matter.
The Hawking 
radiation flux receives non-thermal corrections such that it vanishes 
for late times and the total radiated mass is finite. There are no
static remnants for matter pulses of finite width, 
although a BPP type static remnant appears in
the shock-wave limit. Semi-classical geometries without curvature
singularities can be obtained as well.
Our results indicate that
higher-order loop corrections can remove the singularities encountered
in the one-loop solutions. 

\endtitle

\sect{1. Introduction}

The work on two-dimensional
(2d) dilaton gravity models has shown that they posses the features
one is interested to understand in a realistic gravitational collapse
(for a review see \cite{rev}). 
Especially interesting is the CGHS model \cite{cghs}, which is soluble 
classically and it is a renormalizable 2d field theory. This raises a hope
that the quantum theory may be tractable, so that the properties of quantum 
black holes could be understood. 
As far as the quantum solvability of the CGHS model is concerned, one can find
the physical Hilbert space of states, which is the matter Fock space 
\cite{mik1,mik2,hks,mik4,cjz,mik5}. This exact result allows one to show that 
the quantum theory 
is unitary, because the dynamics is generated by a free-field matter 
Hamiltonian, which can be easily promoted into a Hermitian operator acting
on the
physical Hilbert space \cite{mik4}. Furthermore, one can show that 
evaporating black hole geometries appear in the
semi-classical limit of a unitary gauge quantization of the CGHS model
\cite{mik4,mik5}, which shows
that at least in 2d  evaporating black holes can exist in a unitary quantum
theory. 

Although the results of \cite{mik4,mik5} imply unitary evolution
for the quantum CGHS black hole, 
one would like to see what is the end-state
geometry (i.e. is it a remnant or the black hole completely evaporates with the
information being returned through the Hawking radiation [1]). 
For this one needs the exact effective quantum metric, which can be
obtained either from the exact effective action, or from the expectation 
value of the metric operator. So far only the one-loop perturbative
approximations of the effective metric have been obtained 
\cite{cghs,rst,bpp,mik4}, which are valid in the week-coupling region
of the spacetime. 
Especially interesting is the BPP geometry \cite{bpp}, which describes an
evaporating black hole which ends up as a remnant. This same geometry arises
at the one-loop level of the operator formalism \cite{mik5}, which is 
consistent
with the unitarity. One is then interested to see what is the effect of the
higher loop corrections for the BPP solution, and 
therefore calculating the two-loop corrections is the
simplest thing to do. In this paper we calculate the two loop corrections to
the effective metric by using the operator formalism of \cite{mik5}. The 
advantage of the operator formalism over the conventional effective action
approach is that it gives the spacetime dependence of the effective metric
automatically, while in the effective action approach one has to solve the
effective equations of motion, which in the most cases cannot be done
explicitly. Also at two loops there is a large number of diagrams one would 
have to evaluate in order to obtain the contribution to the effective action.

In section 2 we review the operator formalism of \cite{mik5}. In section 3
we review the BPP solution in the context of the operator formalism, since
it is a starting point for our perturbative calculation. 
In section
4 we calculate the two-loop corrections. In section 5 we examine the two-loop
semi-classical geometry. In section 6 we present our conclusions.

\sect{2. The operator formalism}
  
We start from the classical CGHS action \cite{cghs}
$$ S =  \int_{M} d^2 x \sqrt{-g} \left[ e^{-\f}\left( R + 
 (\nabla \f)^2 + 4\l^2 \right) - \ha (\nabla f)^2 \right]\quad,
\eqno(2.1)$$
where $\f$ is a dilaton scalar field, $f$ is a matter scalar field,  
$g$, $R$ and $\nabla$ are 
determinant, curvature scalar and covariant derivative respectively,
associated with a metric $g_{\m\n}$ on a 2d manifold $M$. The topology of 
$M$ is that of $ {\bf R} \times {\bf R}$. The equations of motion can
be solved in the conformal gauge $ds^2 = -e^{\r}dx^+ dx^-$ as
$$e^{-\r} = e^{-\f} =  - \l^2 x^+ x^- - F_+ - F_- \quad,\quad
f = f_+ (x^+) + f_-(x^-) \quad, \eqno(2.2)$$
where
$$ F_{\pm}= a_{\pm} + b_{\pm}x^{\pm} + 
\int^{x^{\pm}} dy \int^y dz T_{\pm\pm} (z) 
\quad,\eqno(2.3)$$
and $T_{\pm\pm}$ is the matter energy-momentum tensor
$$T_{\pm\pm} = \ha \pa_{\pm} f \pa_{\pm} f \quad.\eqno(2.4)$$ 
The residual conformal invariance can be fixed by a gauge choice $\r = \f$,
and the independent integration constants are $a_+ + a_-$ and $b_{\pm}$.
An equivalent form of the solution (2.2) which is suitable for our purposes,
is given by
$$ F_{\pm}=\a_{\pm} + \b_{\pm}x^{\pm} + \int_{\L^{\pm}}^{x^{\pm}} dy 
(x^{\pm} - y) T_{\pm\pm} (y) 
\quad,\eqno(2.5)$$
where
$$ a_{\pm} = \a_{\pm} + \int^{\L_\pm} dy y T_{\pm\pm} (y) \quad,
\quad b_{\pm} = \b_{\pm} - \int^{\L_\pm} dy T_{\pm\pm}(y) \quad.
\eqno(2.6)$$

It is clear from the solution (2.2) that the independent dynamical
degree of freedom
is a free scalar field $f$. This conclusion also comes out from a reduced
phase space analysis \cite{mik4}. Consequently
the quantum theory is that of a
free mass-less scalar quantum field $f$, propagating on a flat background 
$ds^2_f = -dx^+ dx^-$ with the dilaton and the conformal factor operators 
given by (2.2) \cite{mik4,mik5}. The matter energy-momentum tensor operator is
defined as
$$T_{\pm\pm} = \ha :\pa_{\pm} f \pa_{\pm} f: \quad.\eqno(2.7)$$
The normal ordering in (2.7) is chosen to be with respect to creation
and annihilation operators associated to coordinates $x^\pm$ by
$$f_{\pm} (x^{\pm}) = {1\over\sqrt{2\p}}\int_{0}^{\infty}
{dk\over\sqrt{2\o_k}}\left[ a_{\mp k} e^{-ikx^{\pm}} + a_{\mp k}^{\dagger} 
e^{ikx^{\pm}}\right]\quad,\eqno(2.8)$$
where $\o_k = |k|$. 

The physical Hilbert space of the model is just 
a Fock space ${\cf} (a_k)$ constructed from $a_k^{\dagger}$ acting on
the vacuum $\sket{0}$. The model
is unitary because the dynamics is generated by a free-field
Hamiltonian 
$$H = \int_{-\infty}^{\infty} dk\, \o_k a_{k}^{\dagger} a_k + E_0 \quad,
\eqno(2.9)$$
which is a Hermitian operator acting on ${\cf}$, where
$E_0$ is the vacuum energy. Consequently the states at $t = \ha (x^+ + x^-)=$
const. surfaces are related by a unitary transformation
$$  \Psi (t_2) = e^{-iH(t_2 - t_1)}\Psi(t_1) \quad.\eqno(2.10)$$
One can also define the Heisenberg picture
$$ \Psi_0 = e^{iHt}\Psi(t) \quad,\quad A (t) = e^{iHt}Ae^{-iHt}\quad,
\eqno(2.11)$$
which relates the covariant quantization to the canonical quantization.
For example,
$$f(t,x) = e^{iHt} f(x) e^{-iHt} = {1\over\sqrt{2\p}}\int_{-\infty}^{\infty}
{dk\over\sqrt{2\o_k}}\left[ a_k e^{i(kx-\o_k t)} + a_k^{\dagger} 
e^{-i(kx-\o_k t)}\right]\quad,\eqno(2.12)$$
where $x = \ha (x^+ - x^-)$. Similarly, the operator expressions (2.2) are 
the Heisenberg picture operators.

Given a physical state $\Psi_0$, one can associate an effective metric to 
$\Psi (t)=e^{-iHt}\Psi_0$ as 
$$e^{\f_{eff} (t,x)}
=\sbra{\Psi_0}e^{\f(t,x)}\sket{\Psi_0} \quad,\eqno(2.13)$$ 
where $e^\f$ is the inverse operator of the Heisenberg operator (2.2).
The geometry which is generated by $e^{\f_{eff}}$ via 
$ds^2 = - e^{\f_{eff}}dx^+dx^-$ makes sense only in the regions of $M$
where the quantum fluctuations are small. This will happen if
$$\sqrt{ | \svev{e^{2\f}} - \svev{e^\f}^2  |} << \svev{e^\f}\quad.
\eqno(2.14)$$
The effective conformal factor $e^{\f_{eff}}$ can be calculated perturbatively
by using a series expansion \cite{mik4}
$$(-\l^2 x^+ x^- -  F )^{-1} = e^{\f_0}(1 - e^{\f_0}\d F)^{-1} =
e^{\f_0} \su_{n=0}^{\infty} e^{n\f_0} \d F^n \quad,\eqno(2.15)$$
where $F_0$ is a c-number function, $e^{-\f_0} = -\l^2 x^+ x^- -  F_0 $ and
$\d F = F - F_0$. Then
$$\svev{(-\l^2 x^+ x^- -  F )^{-1}} = 
e^{\f_0} \su_{n=0}^{\infty} e^{n\f_0} \svev{\d F^n}\quad.\eqno(2.16)$$
A convenient choice for $F_0$ is
$$ F_0 = \svev{F_+} + \svev{F_-}\quad,\eqno(2.17) $$
since then the lowest order metric is a one-loop semi-classical metric
$$ e^{-\f_0} = -\l^2 x^+ x^- -  \svev{F_+} - \svev{F_-}\quad.\eqno(2.18) $$
$\Psi_0$ is chosen such that it is as close as possible to a
classical matter distribution $f_0 (x^+)$ describing  a left-moving pulse of
matter. The corresponding classical metric is described by
$$ e^{-\r} = {M(x^+)\over \l} - \l^2 x^+ \D (x^+) - \l^2 x^+ x^-  
\eqno(2.19)$$
where
$$ M (x^+) = \l\int_{-\infty}^{x^+} dy\, y \,T_{++}^0 (y)\quad,\quad
\l^2 \D = \int_{-\infty}^{x^+} dy\, T^0_{++} (y) \quad \eqno(2.20)$$
and $T_{++}^0 = \ha \pa_{+}f_0 \pa_{+} f_0$. The geometry is that of a black 
hole of the mass 
$$M = \lim_{x^+ \to +\infty} M(x^+)\quad,\eqno(2.21)$$ 
and the horizon is at 
$$ x^- = -\D = -\lim_{x^+ \to +\infty} \D (x^+)\quad.\eqno(2.22)$$ 
In the limit of a shock-wave matter distribution, for which
$$T_{++}^0 = a \d (x^+ - x_0^+) \quad,\eqno(2.23)$$
we have
$$ M(x^+) =\l a x_0^+ \theta (x^+ - x_0^+)\quad,\quad 
\D = {a\over \l^2} \quad.\eqno(2.24)$$
The asymptotically 
flat coordinates $(\h^+,\h^-)$ at the past null infinity  are given by
$$ \l x^+ = e^{\l \h^+} \quad,\quad x^- = -\D e^{-\l \h^-} \quad,\eqno(2.25)$$
while the asymptotically flat coordinates $(\s^+,\s^-)$ at the future
null infinity satisfy
$$ \l x^+ = e^{\l \s^+} \quad,\quad \l (x^- + \D ) = - e^{-\l \s^-}\quad.
\eqno(2.26)$$
The corresponding Penrose diagram is given in Fig. 1.
\begin{figure}[h]
\centerline{\psfig{figure=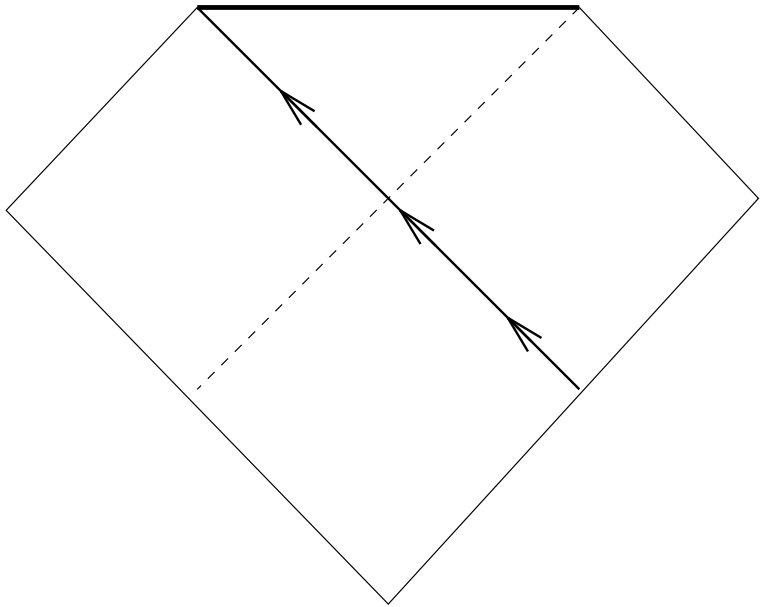}}
\caption{Penrose diagram of the classical CGHS collapse geometry.}
\end{figure}

Note that a change of coordinates $x^\pm \to \x^\pm$ defines a new set of 
creation and annihilation operators through
$$f_\pm = {1\over\sqrt{2\p}}\int_{0}^{\infty}
{dk\over\sqrt{2\o_k}}\left[ b_{\mp k} e^{-ik\x^\pm} + b_{\mp k}^{\dagger} 
e^{ik\x^\pm}\right]\quad.\eqno(2.27)$$
The old and the new creation
and anhilation operators are related by a Bogoluibov transformation
$$ a_k = S^{-1} b_k S = \int_{-\infty}^{\infty} dq ( b_q \a_{qk} + 
b_q^{\dagger} \b_{qk}^* )\quad,\eqno(2.28)$$
and the new vacuum is given by $ \sket{0_\x} = S \sket{0_x}$.
This allows us to
take for $\Psi_0$ a coherent state 
$$\Psi_0 = e^A \sket{0_{\h}^+}\otimes \sket{0_{\h}^-}\quad, \eqno(2.29)$$
where $\sket{0_\h} = \sket{0_\h^+}\otimes\sket{0_\h^-}$ is the vacuum for
the coordinates (2.25), while
$$A = \int_0^{\infty}dk [f_0 (k) a_{-k}^{\dagger} - f_0^{*}(k) a_{-k}] 
\quad,\eqno(2.30)$$ 
where $f_0 (k)$ are the Fourier modes of $f_0 (x^+)$.

\sect{3. One-loop metric}

By using the operator formalism we can calculate perturbatively
the effective metric (2.13) via the expansion (2.16). The lowest
order semi-classical metric will be given by the expression (2.18) which
requires a calculation of the expectation values of the $T_{\pm\pm}$
operators \cite{mik5}. One can show that 
$$\sbra{\Psi_0 } T_{++} \sket{\Psi_0} =
 -{\k\over 4 (x^+)^2} + \ha \left({\pa f_0\over \pa x^+}\right)^2\quad,\quad  
\sbra{\Psi_0} T_{--} \sket{\Psi_0} =  -{\k\over 4 (x^-)^2} \quad,\eqno(3.1)$$
where $\k = {1\over 24 \p}$, so that
$$e^{-\r_0}=e^{-\f_0} = C + b_{\pm}x^\pm - \l^2 x^+ x^- 
- {\k\over 4} \log |\l^2 x^+ x^- | -
 \ha\int_{\L^+}^{x^+} dy^+ (x^+ - y^+)\left({\pa f_0\over \pa
y^+}\right)^2 \quad.\eqno(3.2)$$
The expression (3.2) can also be obtained as a solution 
of the equations of motion of an effective one-loop action \cite{bpp}
$$S_{eff} = S_0 - {\k\over 4}\int d^2 x \sqrt{-g}R\bo^{-1}R - 
\k \int d^2 x \sqrt{-g}(R \f -(\nabla \f )^2 )\quad, \eqno(3.3)$$
where $S_0$ is the CGHS action (2.1).
 
Note that a natural choice for $\L_{\pm}$ is $\L_{\pm} = -\infty$. 
This choice makes
the constant $C$ infinite and $b_{\pm} =0$. We will ignore this infinity,
since from the effective action point of view $C$ is a constant of integration
whose value is determined from a requirement of having a consistent 
semi-classical geometry.
By choosing $C=\frac14 \k [\log (\k/4)-1] $ one can obtain a
consistent  semi-classical geometry \cite{bpp}. In the case of
the shock-wave matter this geometry is well defined in the $x^+
>0,x^- < 0$ quadrant. In the dilaton-vacuum sector ($x^+ < x_0^+$)
the solution (3.2) becomes static
$$e^{-\r_0}= e^{-\f_0} = C - \l^2 x^+ x^- 
- {\k\over 4} \log (-\l^2 x^+ x^- ) \quad,\eqno(3.4)$$ 
and it is defined for $\s \ge \s_{cr}$, where $\s=\log (-\l^2 x^+ x^- )$ is 
the 
static coordinate. At $\s = \s_{cr}$ there is a singularity, and this line is 
interpreted as a boundary of a strong coupling region. This is a common
feature of the semi-classical metrics
\cite{rst} , and a consistent geometry can be defined
for $\s \ge \s_{cr}$ by imposing a reflecting boundary conditions at 
$\s = \s_{cr}$. However, in the operator approach we do not impose reflecting
boundary conditions. A consistent geometry is defined only in the regions of
the spacetime where the metric fluctuations are small, and these coincide with
the weak-coupling region $e^{-\f_0} > 0$ \cite{mik5}.
Note that a curvature singularity occurs at $\s = \s_{cr}$
for $C<\frac14 \k [\log (\k/4)-1] $. Hence this naked singularity will not
appear for $C\ge\frac14 \k [\log (\k/4)-1] $.
 
For $x^+ > x_0^+$ one obtains an evaporating black hole solution
$$e^{-\r_0}=e^{-\f_0} = C + {M\over \l} - \l^2 x^+ (x^- + \D ) 
- {\k\over 4} \log (-\l^2 x^+ x^- ) \quad.\eqno(3.5)$$
The corresponding Hawking radiation flux at the future null-infinity is
determined in the operator formalism by evaluating 
$$ \sbra{\Psi_0} T_{--} (\x^-) \sket{\Psi_0} \quad,\eqno(3.6)$$
where $T_{--}(\x^-)$ is normal ordered with respect to the
asymptotically flat coordinates $\x^{\pm}$  of the metric (3.5) at the 
future null-infinity \cite{mik4,mik5}. The $\x^{\pm}$ coordinates turn out 
to be the same as the ``out"
coordinates (2.26) of the classical black hole solution, so that
$$ 2\p \svev{T_{--}(\x^-)} = {\l^2\over 48}\left[ 1 - (1 + 
\l\D e^{\l\s^-})^{-2}\right]\quad.\eqno(3.7)$$
The expression (3.7) corresponds to a thermal Hawking radiation, with 
$T_H = {\l\over 2\p}$ \cite{{cghs},{gn}}. The Hawking radiation shrinks the 
apparent horizon of the solution (3.5), so that the apparent horizon line 
meets the curvature singularity in a finite proper time, at 
$$ x_i^+ = {1\over \l^2 \D} \left( -\k/4 + e^{1 + {4\over k}(C + M/\l
 )}\right)\quad,\quad
 x_i^- = {- \D\over 1 - {\k\over 4} e^{-1 - {4\over k}(C + M/\l)}}\quad. 
\eqno(3.8)$$
The curvature 
singularity then becomes naked for $x^+ > x^+_i$. However, a static solution 
(3.4) of the form
$$e^{-\r_0}=e^{-\f_0} = {\hat C}  - \l^2 x^+ (x^- + \D ) 
- {\k\over 4} \log (-\l^2 x^+ (x^- + \D)) \eqno(3.9)$$
can be continuously matched to (3.5) along $x^- = x_{i}^-$
if ${\hat C} =\frac14 \k [ \log (\k/4)-1]$. A small
negative energy shock-wave emanates from that point, and for $x^- >
x_{i}^-$ the Hawking radiation stops, while the static geometry (3.9) has
a null ADM mass. There is again a critical line $\tilde{\s} =\tilde{\s}_{cr}$,
corresponding to a singularity of the geometry (3.9). Note that the scalar
curvature of (3.9) is bounded at $x^- = x^-_i$, and the singularity comes from
the pathological behavior of $e^{-\f_0}$, which becomes ill-defined for 
$x^- > x_i^-$. This singularity can be
interpreted as the boundary of the region where higher order
corrections become important. The spatial geometry of the remnant (3.9)
is that of a semi-infinite throat, extending to the strong coupling region.
The Penrose diagram of the one-loop geometry is given in Fig. 2.
\begin{figure}[h]
\centerline{\psfig{figure=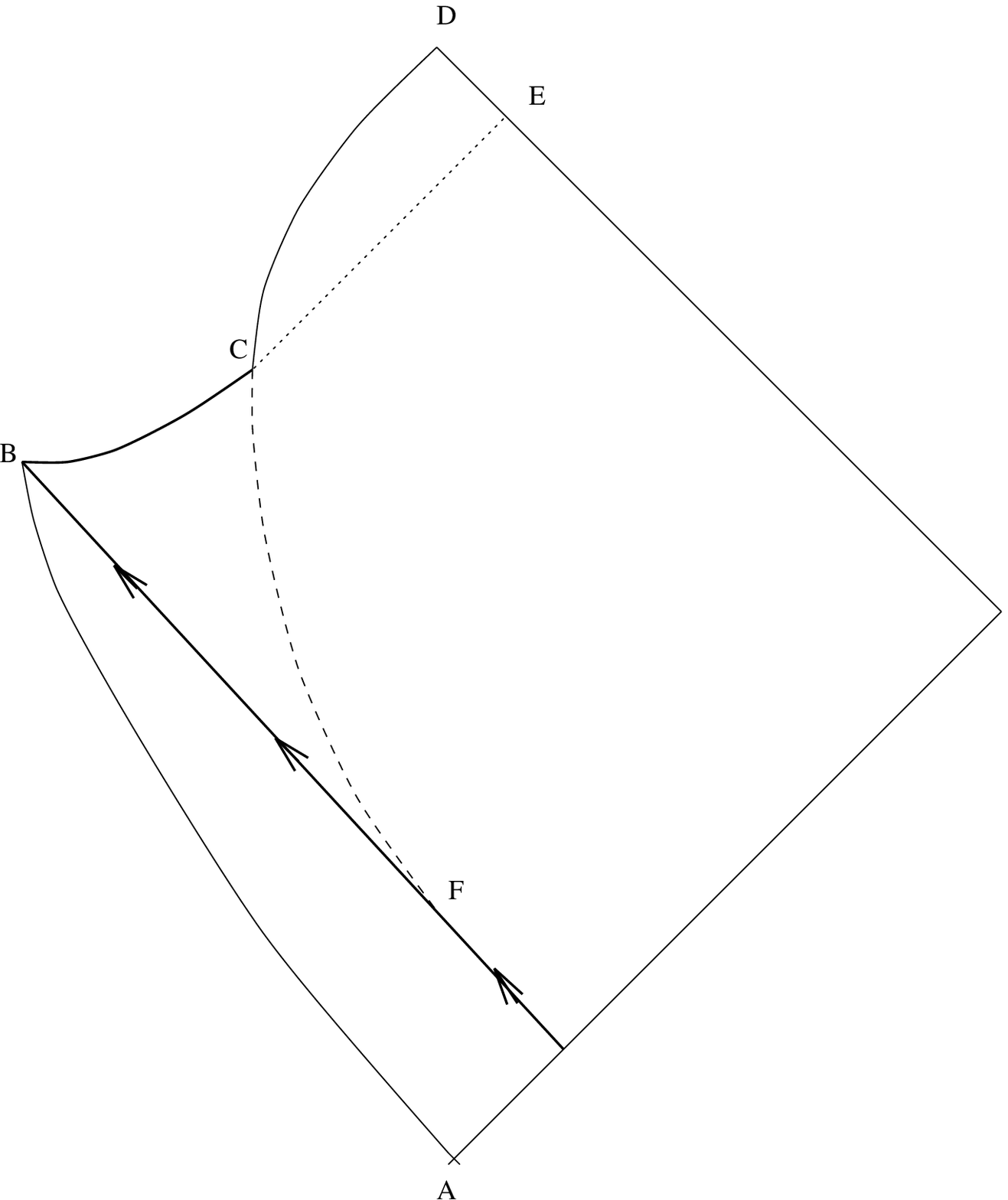}}
\caption{Penrose diagram of the one-loop semi-classical BPP geometry.}
\end{figure}

Note that in the operator approach the spacetime is always ${\bf R}^2$, 
with the geometry of Fig. 2 defined in the region where the metric 
fluctuations and higher order corrections are small. The one-loop 
approximation (2.18) is valid in the regions where
$$ e^{2\f_0}|\svev{\d F^2}| << 1 \quad.\eqno(3.10)$$
The condition (3.10) will certainly break down for $e^{-\f_0}=0$, which also 
determines the position of the curvature singularities of the one-loop
metric. Therefore one can see directly in the operator approach why the
one-loop singularities represent a border of the strong 
coupling region. Also note that (3.10) 
is perturbatively equivalent to  
the condition for small quantum fluctuations (2.14), so that a well-defined 
geometry appears in the region $e^{-\f_0}> 0$, which coincides with the 
weak-coupling region. This situation can be
best described by the Kruskal diagram of Fig. 3.
\begin{figure}[h]
\centerline{\psfig{figure=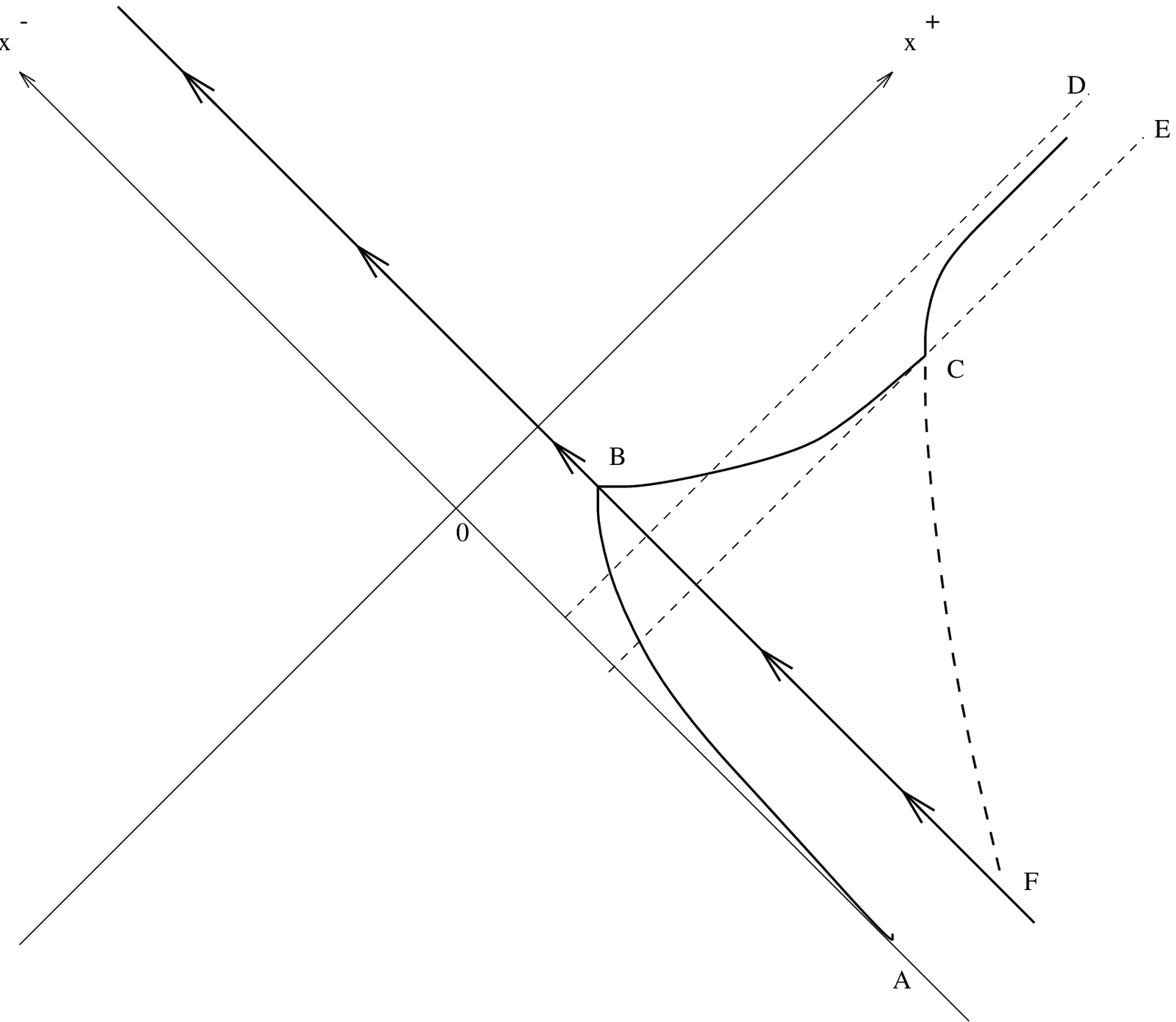}}
\caption{Kruskal diagram of the one-loop semi-classical BPP geometry.}
\end{figure}
The region to the left of the curve ABCD in Fig. 3 corresponds to the 
strong-coupling region.

The operator formalism can also resolve problems connected with a shift
in the classical horizon due to quantum corrections.
From (3.8) it appears as if the classical horizon $x^- = -\D$ has been
shifted to a new position $x^- = x_i^- < -\D$. One can calculate the 
corresponding
Bogoliubov coefficients, and due to this shift, they will be described by
incomplete Gamma functions in the late time approximation \cite{av2}, in
contrast to the
ordinary Gamma functions when there is no shift \cite{gn}. As a result,
the Bogoliubov coefficients will have a different
asymptotic behavior for large frequencies from the standard case, and 
as a consequence the Hawking flux will diverge \cite{av2}.
However, at the one-loop order $\svev{T}$ is given by (3.7), which is finite.
The way out of this paradox is provided by the fact that we are working in a
quantum gravity theory, and therefore the space-time geometry fluctuates. 
Hence the position of the horizon fluctuates, and consistency requires that 
the effective horizon must be at $x^- \ge -\D$.

\sect{4. Two-loop corrections}

In order to calculate the two-loop corrections
it will be useful to redefine $F$ as \cite{mik5} 
$$\li{e^{-\f} =&  \a + \b_{\pm}x^\pm -\l^2 x^+ x^- 
- \int_{\L^\pm}^{x^\pm} dy (x - y)\sbra{0_\h} T_{\pm\pm}(y)\sket{0_\h} \cr -&
 \ha\int_{\L^\pm}^{x^\pm} dy^+ (x^+ - y^+)\left(T_{\pm\pm}(y) - \sbra{0_\h}
T_{\pm\pm}(y) \sket{0_\h} \right)\cr
=& C -\l^2 x^+ x^-  - {\k\over 4} \log |\l^2 x^+ x^- | - F_+ -F_-
\quad, &(4.1)\cr}$$
so that the new $F_{\pm}$ are given as
$$F_{\pm} = \int_{\L^{\pm}}^{x^{\pm}} dy (x^{\pm} - y) \tilde{T}_{\pm\pm} (y) 
\quad, \eqno(4.2)$$
where
$ \tilde{T} = T - \sbra{0_\h} T \sket{0_\h} $
and $\sbra{0_\h} \tilde{T} \sket{0_\h} = 0$. Another convenient
redefinition is to rescale $f(x)$ to $\sqrt{2\p}f(x)$.

The left-moving sector gives corrections due to the matter quantum 
fluctuations \cite{mik4,mik5}. We introduce the following operator ordering
$$:T_{++}(x_1) T_{++}(x_2): = T_{++}(x_1) T_{++}(x_2) -
\sbra{0_\h}T_{++}(x_1) T_{++}(x_2)
\sket{0_\h} \quad,\eqno(4.3)$$
so that
$$\sbra{\Psi_0}:\tilde{T}_{++} (x_1) \tilde{T}_{++} (x_2): \sket{\Psi_0} 
=- {1\over 8\p^2} {\pa f_0\over\pa x_1}{\pa f_0\over\pa x_2} 
\pa_{x_1} \pa_{x_2} \log |\h (x_1) - \h(x_2)|\quad,\eqno(4.4)$$
where 
$$  \sbra{0_{\h}} f (x_1) f (x_2) \sket{0_{\h}}
= -\ha \log |\h^+ (x_1) - \h^+ (x_2 )|
-\ha \log |\h^- (x_1) - \h^- (x_2 )|  \eqno(4.5)$$
has been used \cite{mik5}.

The expression (4.4) is still divergent when $x_1 \to x_2$ since
$$\pa_{x_1} \pa_{x_2} \log |\h^+ (x_1) - \h^+ (x_2)| =
{1\over(x_1 - x_2)^2} + \frac16 D_{x_1}(\h) + O(x_1 - x_2) \quad,
\eqno(4.6)$$
where $D_x (\h) = {\pa_x^3 \h\over \pa_x \h} - \frac32 
\left( {\pa_x^2 \h\over \pa_x \h}\right)^2$ is a Schwarzian derivative. 
Expression (4.6) can be regularized by subtracting 
$(x_1 - x_2)^{-2}$ from $\pa_1 \pa_2 \log(\h_1 - \h_2)$,
which corresponds to using a new ordering \cite{mik5}
$$\li{:T_{++}(x_1) T_{++}(x_2): =& T_{++}(x_1) T_{++}(x_2) -
\sbra{0_\h}T_{++}(x_1) T_{++}(x_2)
\sket{0_\h}\cr -& \sbra{0_x}[A,T_{++}(x_1)][A,T_{++}(x_2)]\sket{0_x} 
\quad.&(4.7)\cr}$$
This ordering gives for $\svev{\d F_+^2}$
$$- {1\over 8\p^2} \prod_{i=1}^2 \int_{\L^+}^{x^+} 
dx_i  (x^+ - x_i )\left( (x_1 x_2 )^{-1} (\log(x_1/x_2))^{-2}
- {1\over (x_1 - x_2 )^2} \right)
{\pa f_0\over\pa x_1}{\pa f_0\over\pa x_2}\,.
 \eqno(4.8)$$

Now we will calculate (4.8) for the shock-wave matter distribution (2.23).
In order to make the calculation simple and well defined,
we take a matter pulse
$$\pa_+ f_0(x^+) = {\sqrt{2a\over\e}}\left[ \theta (x^+ - x_0^+) - 
\theta(x^+ - x_0^+ -\e) \right] \eqno(4.9)$$
where $\e$ is the width of the pulse. When $\e\to 0$ (4.9) gives
$T_{++}$ for the shock-wave. Let $x^+ > x_0^+ + \e$, then $\svev{\d F_+^2}$
is given by
$$ -{a\over 4\e \p^2} \int_{x_0^+}^{x_0^+ + \e}
\int_{x_0^+}^{x_0^+ + \e} dx_1 dx_2 (x^+ - x_1)(x^+ -x_2)
\left({1\over x_1 x_2 (\log (x_1/x_2))^2}-{1\over (x_1 - x_2)^2} \right)
.\eqno(4.10)$$
By using the expansion (4.6), we obtain
$$\li{\svev{\d F_+^2} =& -{a\over 4\e \p^2} \int_{x_0^+}^{x_0^+ + \e}
\int_{x_0^+}^{x_0^+ + \e} dx_1 dx_2 (x^+ - x_1)(x^+ -x_2)
\left({1\over{12 x_1 x_2}} + O(x_1 - x_2) \right)\cr
=&  -{a\over 4\e \p^2}(x^+ - x_0^+)^2  \left( {(\log (1 + \e/x_0^+))^2 
\over 12} + O(\e^4) \right)\quad.&(4.11)\cr} $$
Since $\e$ is small, we have
$$\svev{\d F_+^2}\approx  -{a\e\over 48 \p^2}(x^+ - x_0^+ )^2
\theta(x^+ - x_0^+ )\quad. 
\eqno(4.12)$$
Note that in the limit $\e \to 0$ (4.12) will vanish, so
that there is no two-loop correction in the right-moving
sector for the shock-wave. In the case of arbitrary pulses
$$\svev{\d F_+^2} =  C_+ (x^+)^2 + C_+^{\prime} x^+ + C_+^{\prime\prime}
 \eqno(4.13)$$ 
for $x^+ > x_0^+ + \e$. Note that
$$ C_+ = \int_{x_0^+}^{x_0^+ + \e} \int_{x_0^+}^{x_0^+ + \e} dx_1 dx_2 
\svev{:T_{++}(x_1)T_{++}(x_2):}\quad, \eqno(4.14)$$
in accordance with (4.12).

One can get a similar result by using a $\z$-function 
regularization. One starts from the formula 
$$ \int_0^{\infty} dk\,k\, e^{ik(\h_1 - \h_2 )} = 
-{1\over (\h_1 - \h_2 )^2} \quad,\quad\h_1 - \h_2 \ne 0 \quad,\eqno(4.15)$$
which can be used to rewrite (4.4) as 
$${1\over 8\p^2}\int_0^{\infty} dk\,k\, e^{ik(\h_1 - \h_2 )}
\pa_{x_1}\h\pa_{x_2}\h \pa_{x_1} f_0 \pa_{x_2} f_0 \quad,\eqno(4.16)$$
so that
$$\li{\svev{\d F_+^2} =& {1\over 8\p^2 } \prod_{i=1}^2 
\int_{-\infty}^{\h^+} d\h_i 
(e^{\l(\h^+ - \h_i)} - 1) \int_0^{\infty} dk\,k\, e^{ik(\h_1 -\h_2 )}
{\pa f_0\over\pa \h_1}{\pa f_0\over\pa \h_2}\cr =& 
{1\over 8\p^2}\int_0^{\infty} dk\,k\, |\cf (k, \h^+)|^2 
\quad,&(4.17)\cr}$$
where
$$\cf (k,\h^+) = \int_{-\infty}^{\h^+} d\h e^{ik\h}(e^{\l(\h^+ - \h)} - 1)
{\pa f_0\over\pa \h} \quad.\eqno(4.18)$$
By comparing (4.13) to (4.17), one obtains for the pulse (4.9)
$$C_+ = {a\over 4\p^2 \e}\int_{0}^{\infty} {dk\over k}
\left[ \o^{-2} + \n^{-2} -{2\over \o\n}\cos k (\o - \n)\right]
\eqno(4.19)$$
where $\o = \log \l (x_0^+ + \e)$ and $\n = \log \l x_0^+$. When $\e\to
0$, we get
$$C_+ = {a\over 4 \p^2 }\int_{0}^{\infty} dk 
 k {\e\over\o^{2} (x_0^+)^2} + O(\e^2)\quad.
\eqno(4.20)$$
The infinite integral $\int_0^{\infty} k dk$ can be replaced by a sum 
$\su_{n=1}^{\infty} n$, which can be regularized by using the $\z$ function
$$ \z (s) = \su_{n=1}^{\infty} n^{-s} \quad, \quad
{\rm Re}\, s > 1 \quad, \eqno(4.21)$$
in the following way
$$\li{\int_0^{\infty} k dk =& \lim_{N\to \infty}\int_0^{N} k dk= 
\lim_{N\to \infty}{N^2\over
2}\cr =& \lim_{N\to \infty}\left(\ha N(N+1) - \ha N \right) = 
\lim_{N\to \infty}\left( \su_{n=1}^{N} n -\ha
\su_{n=1}^{N} 1 \right) \cr
=&\su_{n=1}^{\infty} n- \ha \su_{n=1}^{\infty}1  = \z (-1) -\ha \z (0) = 
-{1\over 12} - {1\over 4}\quad. &(4.22)\cr}$$
Therefore
$$ C_+ = -{a\e\over 12 \p^2 x_0^2}\log^{-2} (\l x_0^+)\quad,\eqno(4.23)$$
again a small negative constant, and $\lim_{\e\to 0} C_+ =0$. 
Note that a finite result in (4.23)
has been obtained by analytical continuation of $\z (s)$ to $s = -1$
and $s =0$,
which is the same as if an infinite number has been subtracted from (4.20).
But this is what has been explicitly done in the regularization (4.8), so
that it is no surprise that a similar result was obtained. 

For our purposes it will be instructive to regularize (4.19) when $\e$ is small
but finite by a momentum cutoff. In that case 
$$C_+ = {a\over\p^2 \e\n^2}\int_{0}^{\L} {dk\over k}
\sin^2 {\e k\over 2\l x_0^+} + O(\e) \quad. 
\eqno(4.24)$$
When $\L$ is becoming large, we have
$$C_+ = {a\over 2\p^2 \e\n^2}
\log {\e \g\L\over \l x_0^+} + O(\L^{-1}) \quad, 
\eqno(4.25)$$
where $\g$ is the Euler-Mascheroni constant. Therefore $C_+$ is negative if the
width of the pulse is less then the short distance cutoff $l_c =\l x_0^+/\g\L$
and otherwise $C_+$ is positive.

The right-moving sector determines the
back-reaction of the Hawking radiation \cite{mik4,mik5}. In this case
we use a point-splitting method for regularizing the
operator products. It amounts to calculating an appropriate
limes ($x_{2i-1} \to x_{2i}$) of the expression \cite{mik5}
$$ \sbra{0_{\h}} \prod_{i=1}^{2n} \pa_{x_i^-} f (\h (x_i)) \sket{0_{\h}}
\quad.\eqno(4.26)$$
The expression (4.26) can be calculated by using Wick's theorem and by using
the expression for a two-point function (4.5).
A normal ordering can be defined by an appropriate subtraction of
the products of the terms $(x_i^- - x_j^-)^{-2}$ and 
$\pa_k f_k\cdots \pa_l f_l $ from the expression (4.26)
before taking the limes, such that one obtains a regular expression
after taking the limes. A useful formula for doing this is (4.6).

In the $n=2$ case, one can define 
$$\li{4\p^2 :T_{--}(x_1) T_{--}(x_2): = &\lim_{x_3 \to x_1}\lim_{x_4 \to x_2}
\frac14 \Big(  {\pa f\over\pa x_1}{\pa f\over\pa x_3}{\pa f\over\pa x_2}
{\pa f\over\pa x_4} +\ha
{1\over(x_1 - x_3)^2}{\pa f\over\pa x_2}{\pa f\over\pa x_4}\cr
&+\ha {1\over(x_2 - x_4)^2}{\pa f\over\pa x_1}{\pa f\over\pa x_3}
+ \frac14 {1\over(x_1 - x_3)^2}{1\over(x_2 - x_4)^2}
 \Big) \,&(4.27)\cr}$$
where $x_i$ denotes $x_i^-$. This ordering gives
$$\li{ \sbra{0_{\h}}: T_{--}(x_1) T_{--} (x_2): \sket{0_{\h}}
=&{1\over 32\p^2}\left( \pa_{x_1} \pa_{x_2} 
\log |\h^- (x_1) - \h^- (x_2)|  \right)^2\cr 
+&\svev{T_{--} (x_1)}\svev{T_{--} (x_2)}\quad.&(4.28)\cr}$$
The expression (4.28) is still divergent when $x_1 \to x_2$, and it will be
the source of divergence in
$$\svev{\d F_-^2}= {1\over 32\p^2}\int_{\L^-}^{x^-}\int_{\L^-}^{x^-} 
dx_1 dx_2  (x^- - x_1 )(x^- - x_2 )\left( \pa_{x_1} \pa_{x_2} 
\log |\h^- (x_1) - \h^- (x_2)| \right)^2 \, .
\eqno(4.29)$$

One way to regularize (4.28) is by changing the definition (4.27)
\cite{mik4,mik5}. This amounts to
$$\li{ \sbra{0_{\h}}: T_{--}(x_1) T_{--} (x_2): \sket{0_{\h}}
=&{1\over 32\p^2}\left( \pa_{x_1} \pa_{x_2} \log |\h^- (x_1) - \h^- (x_2)|  
- {1\over (x_1 - x_2)^2}\right)^2\cr 
+&\svev{T_{--} (x_1)}\svev{T_{--} (x_2)}\quad,&(4.30)\cr}$$
which is finite for $x_1 \to x_2$ due to (4.6). Consequently
$\svev{\d F_-^2}$ is given by
$$ {1\over 32\p^2} \prod_{i=1}^2 \int_{\L^-}^{x^-} 
dx_i  (x^- - x_i )\left( \pa_{x_1} \pa_{x_2} 
\log |\h^- (x_1) - \h^- (x_2)|- {1\over (x_1 - x_2 )^2} \right)^2\,, 
\eqno(4.31)$$
which is finite for $\L_-$ finite. However, it is natural to take $\L_-
= - \infty$. In that case the integral (4.31) is independent of $x^-$,
which can be seen by making a substitution 
$$ x_i = x^- e^{ t_i} \eqno(4.32)$$
so that
$$\svev{\d F_-^2}= {1\over 32\p^2}\int_{0}^{\infty}\int_{0}^{\infty} 
dt_1 dt_2 e^{t_1 + t_2} (1- e^{t_1} )(1- e^{t_2} )
\left( e^{-{(t_1 + t_2)\over (t_1 - t_2 )^2}} - (e^{t_1} -
e^{t_2} )^{-2}\right)^2 \, .
\eqno(4.33)$$
The integral (4.33)
is a divergent constant, and the way it arises
is similar to the one-loop constant $C$. Hence we will consider (4.33) as a 
finite constant, whose value is determined from some self-consistency 
requirement, and therefore
$$\svev{\d F_-^2} = C_- \quad.\eqno(4.34)$$

Note that the result (4.34) can be also obtained with the $\z$-function 
regularization. Namely, by making the substitution (4.32) 
in the integral (4.29) we obtain
$$\svev{\d F_-^2}= {1\over 32\p^2}\int_{0}^{\infty}\int_{0}^{\infty} 
dt_1 dt_2  (1- e^{-t_1} )(1- e^{-t_2} )(t_1 - t_2 )^{-4} \, .
\eqno(4.35)$$
By repeated differentiation of (4.14) with respect to $\h$ we obtain
$$ {1\over (t_1 - t_2 )^4} =\frac16 \int_0^{\infty} dk\,k^3\, 
e^{ik(t_1 - t_2 )}  \quad,\eqno(4.36)$$
so that
$$C_- =\lim_{l \to \infty} {1\over 6\cdot 32\p^2}\int_{0}^{\infty} 
dk k^3 |J_l (k)|^2
\eqno(4.37)$$
where
$$ J_l (k) = \int_0^l dt (1- e^{-t}) e^{ikt} \quad.\eqno(4.38)$$
For large $l$ we have
$$ J_l (k) = { e^{ikl}-1\over ik} + {1\over ik - 1} + O(e^{-l})\eqno(4.39) $$
so that
$$\li{ C_- =&\lim_{l \to \infty}{1\over 6\cdot 32\p^2}\int_{0}^{\infty} 
dk k^3 \left( {k^2 + 2\over k^2 ( 1 + k^2)} - {2\sin kl\over k(1+k^2)}
 - {2 \cos kl\over k^2 (1+k^2)} \right)\cr
=&\lim_{l \to \infty}{1\over 6\cdot 32\p^2}\left( \int_{0}^{\infty} 
dk k {k^2 + 2\over k^2 + 1} - {2\over l} \int_{0}^{\infty} dx 
{x^2 \sin x \over l^2 +x^2 }
 -\int_{0}^{\infty}dx  {2x \cos x \over l^2 + x^2 } \right)\cr
=&{1\over 6\cdot 32\p^2}\int_{0}^{\infty} 
dk k \left( 1 + {1\over k^2 + 1}\right)\quad.&(4.40)\cr}$$

The last integral in (4.40) can be rewritten as 
$$\li{ I =& \lim_{N\to\infty}\left( \int_{0}^{N} kdk +\ha\int_{1}^{\infty} 
{dx \over x}\exp (-e^{-N} x) \right)\cr
=& \lim_{N\to\infty}[ \ha N(N+1) - \ha \g ] 
= \su_{1}^{\infty} n - \ha\g = \z (-1) - \ha \g\quad,&(4.41)\cr} $$
where $\g$ is the Euler-Mascheroni constant and we used
$$  \int_1^{\infty} e^{-\e x} {dx\over x} =-{\rm Ei} (-\e) = - \g - 
\log\e + O(\e) \quad.\eqno(4.42)$$ 
Therefore we obtain
$$C_- =-{1\over 12\cdot 32\p^2}\left({1\over 6} + \g \right)\quad, 
\eqno(4.43)$$
a negative constant. 

\sect{5. Two-loop semi-classical geometry}

The analysis in the previous section implies that
$$ \svev{\d F^2} = C_- + C_+ (x^+ - x^+_0 )^2 \theta (x^+ - x^+_0)\quad,
 \eqno(5.1)$$
for narrow matter pulses. We have also shown that a regularization exists
such that $C_-$ is a negative constant, while $C_+$ depends on the
shape of the matter pulse. Clearly the constants $C_{\pm}$
are regularization dependent. This regularization dependence is not a problem,
since from the effective action approach point of view $C_{\pm}$ can be
considered as integrals of motion 
(the 
two-loop equations of motion will be of higher order in spacetime
derivatives, and therefore new integration constants will appear). Therefore 
the range of $C_{\pm}$
could be determined from some appropriate consistency condition.
Also, in the following analysis we will show
that the corresponding geometries are essentially determined by the sign of 
$C_{+}$. Note that in the case of arbitrary pulses of compact support 
$\svev{\d F^2}$ is quadratic in $x^+$ for $x^+$ outside of the support interval
(see (4.13)), so that (5.1) is a good effective approximation in the
general case. However, $C_+$ can be then of both signs, positive or negative,
depending on the matter pulse profile, which can be seen from (4.14) and
(4.25). 

The two-loop metric can be written as 
$$ds^2 = - e^{\f_2}dx^+dx^-\quad;$$
$$ e^{\f_2} = e^{\f_0} \left[ 1 + e^{2\f_0}(C_- + C_+ (x^+ - x_0^+)^2
\theta (x^+ - x_0^+)) \right]\quad, \eqno(5.2)$$
where $e^{\f_0}$ is the one-loop dilaton solution
$$e^{-\f_0} = C - a (x^+ - x_0^+) \theta (x^+ - x_0^+) - \l^2 x^+ x^- 
-\frac{\k}4 \log |\l^2 x^+ x^-| \quad.\eqno(5.3)$$
We introduce new constants $\a$ and $\b$ such that $C_- = - \a^2$
and $C_+ = \pm \b^2$. 
As in the one-loop case, the relevant quadrant is
$x^+ \ge 0,x^- \le 0$. For
 $x^+ < x_0^+$ the solution (5.2) is static, and
it can be written as
$$ e^{\f_2} = e^{\f_0} \left[ 1 - e^{2\f_0}\a^2 \right]\quad,\quad
e^{-\f_0} = C + e^\s -\frac{\k}4 \s \quad,\eqno(5.4)$$
where $\s =\log (-\l^2 x^+ x^-)$ is the static coordinate. The solution (5.4)
describes a two-loop corrected dilaton vacuum. 
The corresponding scalar curvature 
will diverge on the curve
$$ e^{-2\f_0} - \a^2 = 0 \quad.\eqno(5.5)$$
The equation (5.5) has two solutions $e^{-\f_0} =\pm \a$, which
correspond to the one-loop singularity lines where $C$ is replaced by 
$C \mp \a$. Since the shock-wave intersects first the $C-\a$ line ($\a >0$),
this will be the critical line. The semi-classical geometry will be defined
for $\s \ge \s_c$, where $C + e^{\s_c} -\frac{\k}4 \s_c = \a$. The curvature
singularity will be absent for 
$C\ge \a -\frac{k}4 ( 1- \log\frac{k}4)$.
Therefore if we want to avoid a naked singularity at two loops, the
BPP choice of the one-loop $C$ has to be increased to 
$$C= \a + \frac{k}4( \log\frac{k}4 -1)\quad.\eqno(5.6)$$ 
Note that in the quadrants $x^+ x^- \ge 0$ naked singularities are present
for any value of $C$. However, for the observer located at the right spatial 
infinity, these are located in the strong coupling region
$e^{-\f_0} \le 0$, where the two-loop approximation breaks down, 
and therefore can be ignored. Also note that for $C_- = +\a^2$ no curvature
singularities appear, and the only singularity comes from $\f$ becoming
complex for $e^{-\f_0}<0$.

For $x^+ > x_0^+$ the solution (5.2) becomes 
$$ e^{\f_2} = e^{\f_0} \left[ 1 + e^{2\f_0}(-\a^2 + C_+ (x^+ - x_0^+)^2)
 \right]\quad, \eqno(5.7)$$
where
$$e^{-\f_0} = C + {M\over\l} - \l^2 x^+ (x^- + \D) 
-\frac{\k}4 \log(-\l^2 x^+ x^-)\quad.\eqno(5.8)$$
It describes a two-loop corrected evaporating
black hole geometry. The curvature singularity is given by the curve
$$ e^{-2\f_0} - \a^2 +C_+ (x^+ - x_0^+)^2 = 0 \quad.\eqno(5.9)$$
The curve (5.9) can be parameterized as
$$\li{ x^+ x^- =& - e^\s \cr
       x^+ - x_0^+ =& {\D C_\s \pm \sqrt{\a^2(\D^2 + C_+ ) - C_+ C_\s^2}
 \over \D^2 + C_+} &(5.10)\cr}$$
where $C_\s = C + e^\s -\frac{k}4 \s$, and we have set $\l =1$. 
When $C_+ = -\b^2 $, the relevant branch of
(5.10) is a curve which starts from $(x^+_0, -k/4x^+_0)$ and it goes
to $x^+ =\infty$ with an asymptote $x^- = -\D -\b$. When $C_+ =\b^2$, (5.9)
is a closed curve, which starts from $(x^+_0, -k/4x^+_0)$, extends to
$x^+ = x^+_0 + (\a\D/\b)(\D^2 + \b^2 )^{-\ha}$, where it turns back and ends 
up at $x^+ = x^+_0$ line. 
  
The apparent horizon curve is given by the equation
$\pa_+ \f_2 = 0$, which can be rewritten as
$$  x^- + \D +{k\over 4\l^2 x^+} = -{ 2C_+ (x^+ - x_0^+)e^{-\f_0}/\l^2\over
e^{-2\f_0} - 3\a^2 + 3 C_+ (x^+ - x_0^+)^2 } \quad.\eqno(5.11)$$
When $C_+ <0$, this curve starts from $(x^+_0, -\D -k/4x^+_0)$, intersects 
$x^- =-\D -\b$, and then asymptotically approaches this line as 
$x^+ \to\infty$.
This behavior implies that the apparent horizon  must intersect the curvature
singularity at some $(x^+_i, x^-_i)$, so that for $x^+ > x^+_i$ there
is a naked singularity. If $C_+ >0$, then (5.11) is a closed curve, which
starts from $(x^+_0, -\D -k/4x^+_0)$, extends to $x^+ \approx x^+_0 + 5/4\b$,
and it comes back to $x^+ = x^+_0$ line. The naked singularity will
appear in this case, 
but it will be located in the strong coupling region $e^{-\f_0} \le 0$, 
and therefore it can be ignored. Note that this does not happen in the $C_+ <0$
case, where the naked singularity lies outside the strong coupling region
(see Figs 4 and 5). As we are going to see later, this behavior is 
correlated with the behavior of the Hawking flux.

\begin{figure}[h]
\centerline{\psfig{figure=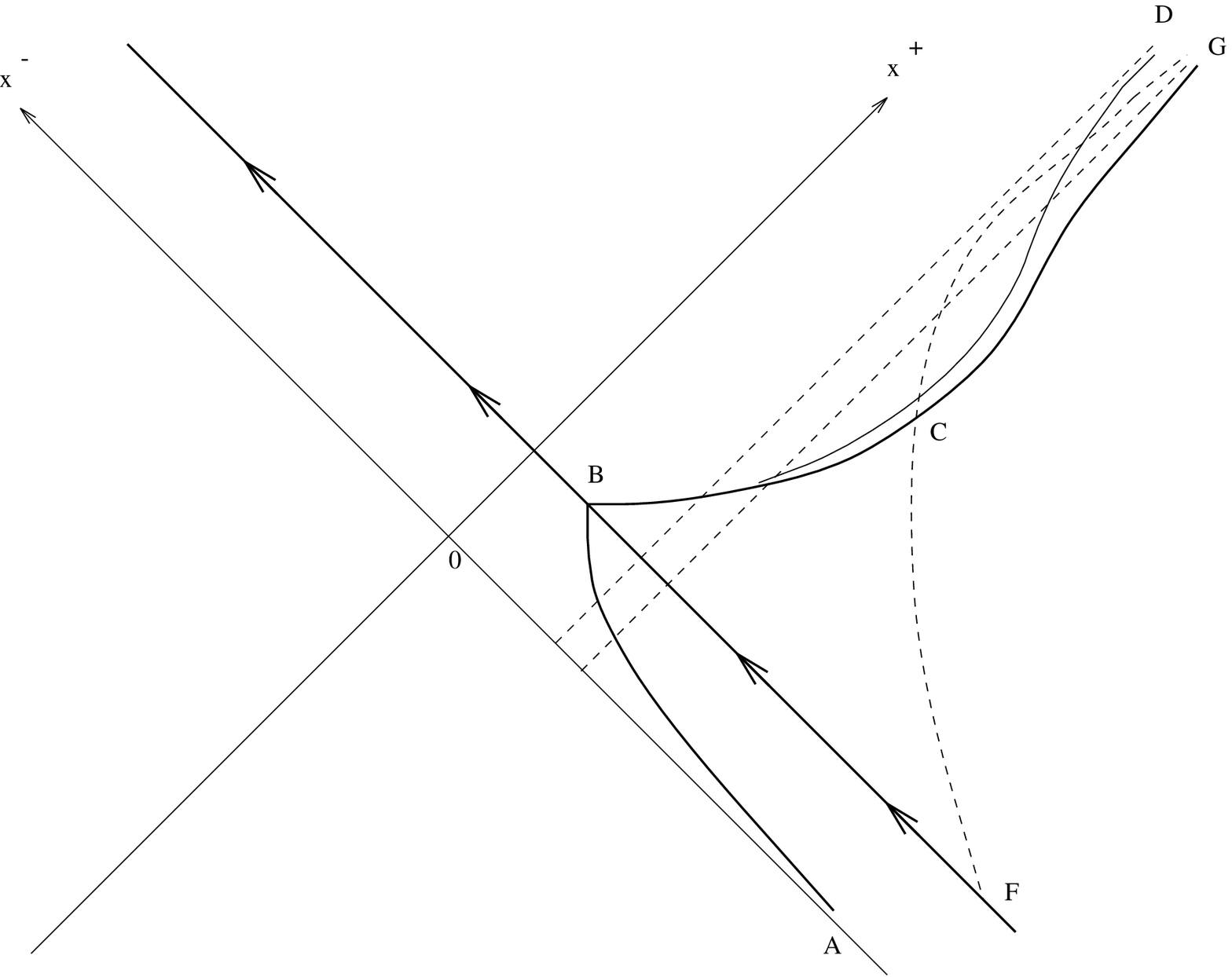}}
\caption{Kruskal diagram of the two-loop semi-classical geometry when 
$C_+ <0$. The straight dashed lines passing through D and G are $x^- = -\D$ and
$x^- = -\D -\b$ lines, respectively. The curve BD
is the strong-coupling border $e^{-\f_0} =0$, and the apparent horizon curve
FCG intersects the curvature singularity curve BCG in the weak-coupling 
semi-classical region $e^{-\f_0} > 0$.}
\end{figure}

\begin{figure}[h]
\centerline{\psfig{figure=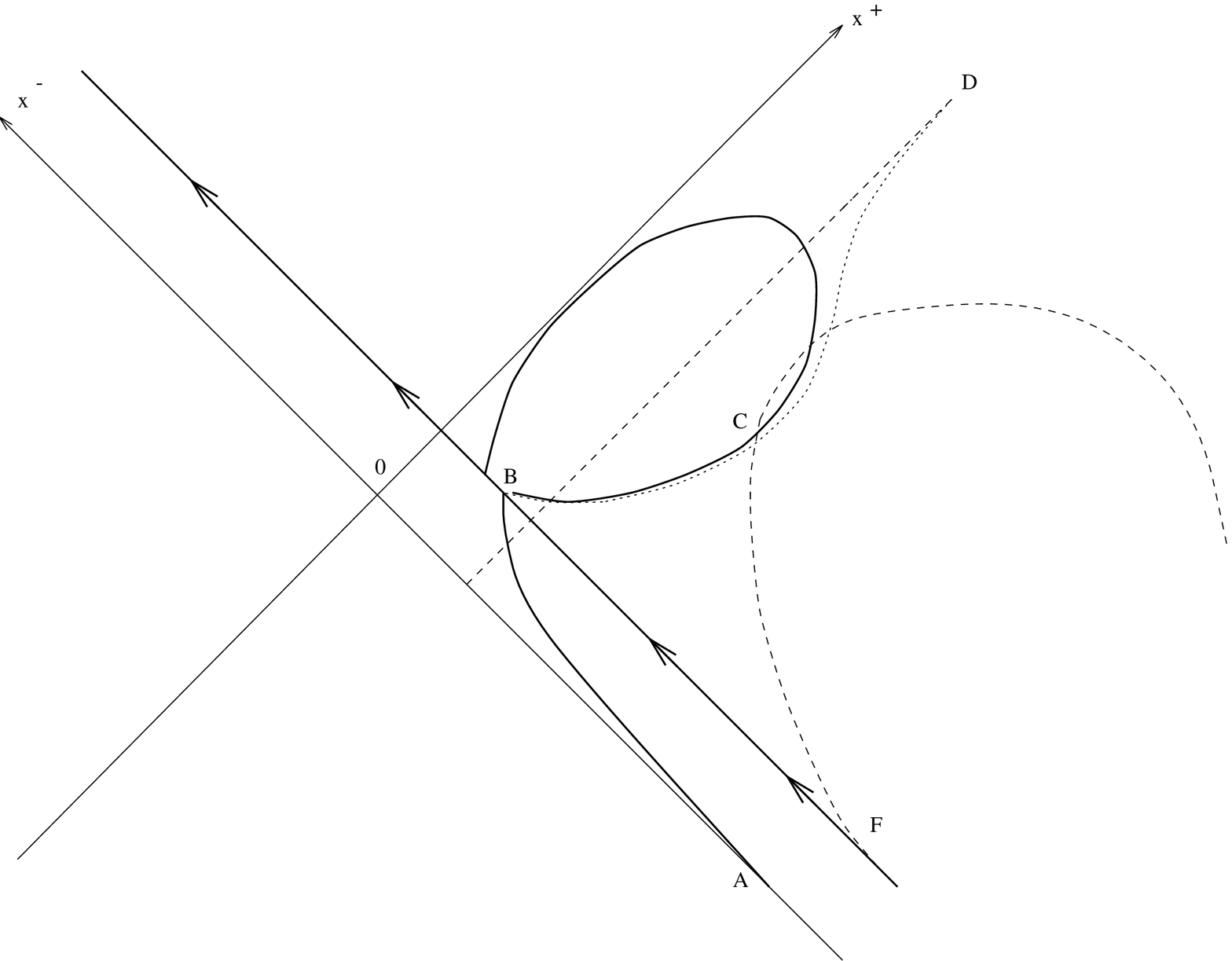}}
\caption{Kruskal diagram of the two-loop semi-classical geometry when 
$C_+ >0$. The apparent horizon curve FC intersects the curvature singularity
curve BC inside the strong-coupling region, bordered by the curve BD.}
\end{figure}

When $\b = 0$, the equation (5.9) becomes a shock wave 
singularity equation (5.5), and
the relevant root of that equation is $e^{-\f_0} - \a = 0$. This solution 
describes
a singularity line of the one-loop metric with a smaller ADM mass
$ C + {M\over \l} - \a$. For a minimal allowed value of $C$ (5.6), this
becomes the one-loop metric with $C=\frac{k}4 (\log\frac{k}4 -1)$.
For a very small or vanishing $\b$,
the $\b$ terms can be neglected in (5.11) and
one obtains the one-loop apparent
horizon line equation. Therefore the intersection point of the apparent horizon
line and the curvature singularity line for the shock-wave at two-loops
is given by the one-loop expressions (3.8). 
The change in the intersection point (3.8) when $\b \ne 0$ can be 
evaluated perturbatively, and it is given by
$$\li{ \d x_i^+ =& -{C_+\over 2\a} {(x_i^+ - x_0^+)^2\over ax_i^+} x_i^- 
+ O(C_+^2)\cr
 \d x_i^- =& -{2\l^2 C_+\over \k\a} {(x_i^+ - x_0^+)^2\over a} x_i^-
x_i^+ + O(C_+^2) \quad.&(5.12)\cr}$$

Therefore when $C_+ <0$ a naked singularity will appear
in the region $x^- > x_i^- , x^+ > x_i^+$, unless we
impose an
appropriate boundary condition. In the shock-wave case ($C_+ =0$), one can
impose a static solution (5.4) for $x^- > x_i^-$ 
$$e^{\f_2} = e^{\f_0} \left[ 1 - e^{2\f_0}\a^2 \right]\quad,\quad
e^{-\f_0} = {\hat C}  - \l^2 x^+ (x^- + \D) 
-\frac{\k}4 \log(-\l^2 x^+ (x^- + \D))\quad,\eqno(5.13)$$
which can be continuously matched to (5.7) at $x^- =x_i^-$ if 
$${\hat C}= \a -{\k\over 4}(1-\log{\k\over 4})  \quad.\eqno(5.14)$$
The metric (5.13) with the value of $\hat{C}$ given by (5.14)
does not have a curvature
singularity at $\s =\frac1{\l}\log(-\l^2 x^+ (x^- + \D))= \s_{cr}$, and the 
naked singularity is removed. In this 
case $\s =\s_{cr}$ corresponds to a curve where $e^{-\f_0}$ is
ill-defined (branch-point singularity), in a complete analogy with the 
one-loop case. However,
when $C_+ = -\b^2 \ne 0$, there is no static 
two-loop dilaton vacuum solution which can be continuously matched to (5.7).
The best one can do is to take
$$ e^{\f_2} = e^{\f_0} \left[ 1 - e^{2\f_0}(\a^2 + \b^2 (x^+ - x_0^+)^2)
 \right]\quad, \eqno(5.15)$$
for $x^- \ge x_i^-$ and $x^+ \ge x^+_i$, where $e^{\f_0}$ is given by (5.13) 
and (5.14). Still, the naked singularity remains. 
Hence the two-loop corrections make the one-loop remnant unstable. Note that
when $C_+ = \b^2$, there is no curvature singularity for 
$x^+ > x^+_0 + (\a\D/\b)(\D^2 + \b^2 )^{-\ha}$, and 
therefore in this case there is no need for a sewing procedure. By examining
the Hawking flux, we will see that this solution has a well defined flux of 
emitted particles at future null infinity.
  
In order to calculate the Hawking flux we need
the asymptotically flat coordinates $(\s^+,\s^-)$ at $\ci^+_R$. These are
given by
$$ \l x^+ = e^{\l\s^+} \quad,\quad \l(x^- + \D ) = - e^{-\l \tilde{\s}}
\quad,\eqno(5.16)$$
where
$$ \tilde{\s} = \s^- - {C_+\over 2\l^3} e^{2\l\tilde{\s}}\quad. \eqno(5.17)$$
This can be written as
$$\s^- = -{1\over \l} \log [-\l(\D + x^-)] + 
{C_+\over 2\l^5} (\D + x^-)^{-2} \quad,\eqno(5.18)$$
so that a non-zero correction appears at two loops. When $C_+ < 0$, the 
coordinate
change (5.16) is well-defined for $x^- < -\D - {\b\over\l^2}$,
otherwise it is two-valued. When $C_+ >  0$, then this coordinate
singularity is absent.
A direct consequence of (5.18) is that
the Hawking flux at $\ci^+$ will not have the thermal form (3.7), which can be 
seen by evaluating (3.6) for $\x^- = \s^-$. This gives  
$$ 2\p \svev{T_{--}}|_{\ci^+} = -{1\over 24}\left(
{\h^{\prime\prime\prime} \over\h^{\prime}} - 
\frac32 \left({\h^{\prime\prime}\over\h^{\prime}}\right)^2\right) 
\quad,\eqno(5.19) $$
where the primes stand for derivatives with respect to $\s^-$ and 
$x^- = -\l\D e^{-\l\h}$. One then obtains
$$ 2\p\svev{T_{--}}|_{\ci^+} = T(y)= {1\over 24}
{y^4 [ y^4 (-y\D + \ha\D^2 ) - C_+ P_4 (y)
+ C_+^2 P_2 (y)]\over (y-\D)^2 (C_+ + y^2)^4}\quad,\eqno(5.20) $$
where $y = x^- + \D$, 
$$P_2 (y)= -2y^2 + 3\D y -\frac32 \D^2 \quad,\quad 
P_4 (y) = y^2 ( -4y^2 + 10 \D y - 5\D^2 )\quad,  \eqno(5.21)$$
and we have set $\l =1$. In Fig. 6  we give plots of $T(y)$ versus $y$ 
for $C_+ <0$ and $C_+ >0$, respectively.
\begin{figure}[h]
\centerline{\psfig{figure=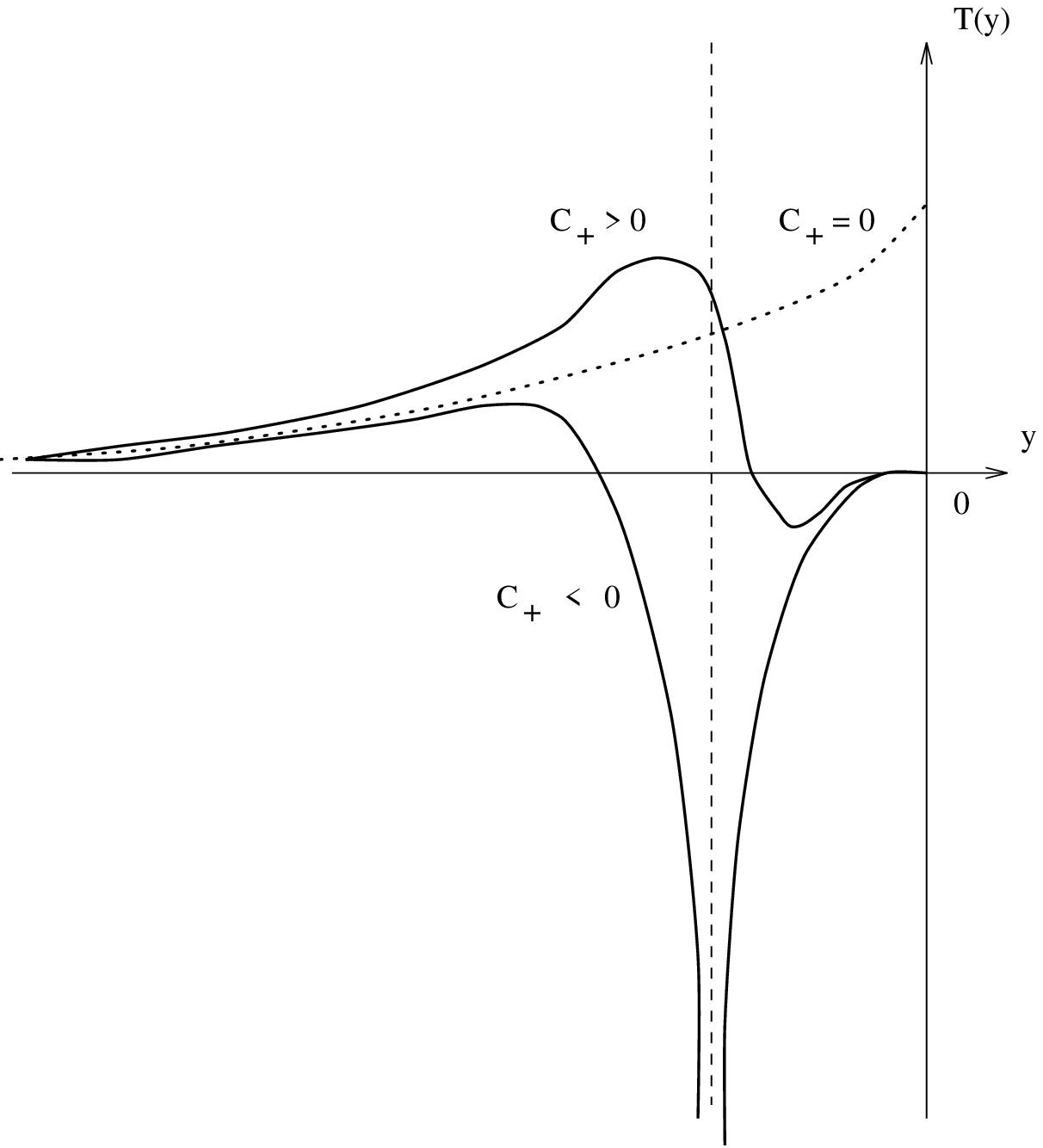}}
\caption{Plots of the two-loop Hawking flux for $C_+ <0$, $C_+ =0$ and
$C_+ > 0$. The vertical dashed line is $y=-\b$ line.}
\end{figure} 
The dotted line represents the classical (zero loops) Hawking 
flux
$$ T_0 (y) = {1\over 24}{\ha\D^2 - y\D \over (y-\D)^2 }\quad. \eqno(5.22)$$ 

When $C_+ <0$, $T(y)$ is close to $T_0 (y)$ for early times, but as
$y \to 0 $, i.e. for late times, $T(y)$ goes to zero and then 
it diverges to $-\infty$ at $y = -\b$. Note that in this case there is a naked
singularity in the weak coupling region. It has been also found at the one-loop
order that naked singularities are accompanied by pathological Hawking
fluxes \cite{vw}, and it has been argued there
that such a catastrophic divergence of the flux is a generic feature whenever
naked singularities occur. On the other hand, when $C_+ >0$, the naked 
singularity is absent, and $T(y)$ is finite and continuous in the
semi-classical region. Moreover, it smoothly
goes to zero for late times, so that the total radiated mass is finite
$$M_{rad}= \int_{-\infty}^{0} dy{d\s^-\over dy} T(y)< \infty \quad,
\eqno(5.23)$$
which are features not present at zero and one loop order. Actually, (5.23) is
finite for the BPP solution, but this is done somewhat artificially, by  
taking (5.22) to be zero for $y > x_i^- + \D$. As a result of this 
discontinuous change in the flux, a shock-wave of negative energy emanates
from $(x_i^+, x_i^-)$ (a thunder-pop), so that the total energy is conserved. 
The two-loop solution with $C_+ >0$ has a desired
property that the Hawking radiation turns off itself for late times, but then
becomes negative, and it goes to zero at $y=0$ (see Fig. 6). The appearance
of this negative energy flux is a characteristic of a situation where all
of the infalling matter gets out, since then from the energy conservation it 
follows
that $M_{rad} =0$, which can be only achieved if $T(y)$ becomes negative for 
late times. This behavior has been observed in the case of the one-loop BPP
solution with reflecting boundary conditions, when $M < M_{cr}$ so that a
black hole is not formed, and hence all of the infalling matter gets out 
(it appears at $x^+ = +\infty$, the right future null-infinity)\cite{bpp2}. 
In our case, the black hole forms and there are no reflecting boundary 
conditions, so that the infalling matter will reach the left future 
null-infinity 
($x^- = +\infty$). This can be interpreted as matter getting out, provided
there are no singularities at $x^- = +\infty$. However, the line 
$x^- = +\infty$ lays in
the strong-coupling region, and without the knowledge of the full 
non-perturbative
solution we can not say what exactly happens there. Still,
all this indicates that the Hawking flux will
deviate significantly from thermality for very late times. Also 
the behavior of $T(y)$ is consistent with the creation of 
particle-antiparticle pairs, where particles reaching infinity give rise to 
positive $T(y)$, while antiparticles carrying negative energy give rise to 
negative $T(y)$ \cite{bpp2}. 

Note that (5.23)
is only a function of $\D$, $C_+ =\b^2$ and $\l$, and it does not depend on 
the 
incoming mass $M$. Also $M_{rad}(\b ,\D)$ diverges for small $\b$'s, so that 
for a sufficiently small $\b$ one will have $M_{rad}> M$, and these solutions
will violate the energy conservation. 
This can only mean that the 
higher order corrections have not been taken into account. On the other hand, 
if one wants to have two-loop solutions which are consistent with the
energy conservation, $\b$ should be larger (which means wider matter pulses,
see (4.25)) so that $ M_{rad}(\b,\D) < M$. Solutions with $M_{rad} >0$ are
consistent with the appearance of a remnant, while the $M_{rad} =0$ solution
is consistent with the complete evaporation of the black hole. When $\b =0$,
then one has a BPP type solution, with a massless remnant. In that case
$$M_{rad} = M -k{\l\D\over 4x_i^- } > M \quad,\eqno(5.24)$$ 
and the energy conservation is insured by emission of a thunder-pop, 
of negative energy $ k{\l\D\over 4x_i^- }$, which 
appears because the remnant metric and the black hole metric are not sewn up
smoothly.

\sect{6. Conclusions}

The two-loop corrections to the effective metric take a simple form (5.1),
which is valid for arbitrary matter pulses when
$x^+$ is outside of the pulse. The constants $C_\pm$ are 
regularization dependent, and one can consider them as integrals of motion.
The corresponding semi-classical geometries are
essentially determined by the sign of $C_+$. When $C_+ \ge 0$, consistent
semi-classical geometries appear, where consistency means absence of naked
singularities in the regions of the spacetime where the semi-classical
approximation is valid and the total radiated mass is less then the
infalling mass.
The solutions with $C_+ =0$ (shock-wave geometry) are special,
in the sense that their geometry is essentially the same as the one-loop
geometry, with the two-loop corrected static remnant appearing as the 
end-state. The Hawking flux is exactly the same as the one-loop flux, and a
thunder-pop appears. However, when $C_+ >0$, i.e. when the pulse has a 
reasonable 
width, the Hawking flux receives two-loop non-thermal corrections, and it
becomes a continuous function of time. There is no remnant in the weak-coupling
region, and the Hawking flux becomes negative for very late times (approaching
zero) indicating particle-antiparticle pair creation. 
For sufficiently wide pulses one can have $M_{rad} < M$, so that the total 
energy is 
conserved, in agreement with the unitarity. Solutions with $M_{rad}>0$ are
consistent with a remnant appearing in the strong-coupling region. However,
in order to check this, one would need the full nonperturbative solution,
especially because $M_{rad}$ could then vanish, implying that all infalling 
matter gets out and the black hole completely evaporates. Note that appearance
of remnants is somewhat unnatural in models where infalling matter does not 
couple to
the gravitational field, and in order to insure the energy conservation one
has to prevent the infalling matter to reach the future infinity. Apart
from simply terminating the evolution by hand after some time $x^-_0$ 
(as is done
in the BPP case), a more natural solution would be that a singularity in
geometry appears at $x^- = +\infty$.   
Also note that when $C_+ >0$ and $C_- >0$ no curvature 
singularities appear at all, although singularities in $\f$ remain
($\f$ becomes complex in the strong-coupling region).
Whether this can be interpreted in favor of the no-remnant
scenario it is difficult to say, since at this moment we do not know what
happens in the strong coupling region $e^{-\f_0}\le 0$. Still, it is very
indicative that the two-loop corrections can cure the problems encountered
in the lower loop approximations. This gives us a hope that the full
non-perturbative solution will be
well defined in the semi-classical regions of the spacetime (i.e. regions
where the metric fluctuations are small), so that one can obtain a definite
answer about the fate of the 2d black hole. 

Note that the back-reaction effect on the Hawking radiation flux is such
that the Hawking radiation must deviate significantly from thermality in the
late stages of evaporation. This behavior is expected from general arguments,
and it is very encouraging that it is recovered in a concrete model. An 
analysis of the Bogolibov coefficients confirms this \cite{av2}, and it will
be interesting to better understand the effect of the horizon fluctuations,
which is described by the operator (2.22).

When $C_- <0$, a naked singularity can appear in
the dilaton vacuum sector if the one-loop constant 
$C <\a +\frac{k}4(\log\frac{k}4 + 1) $. However, this  singularity does not
affect the Hawking flux since it is independent of $C_-$ and $C$.
Also the constant $C$ does not depend on the incoming matter state,
and hence it can be chosen freely, such that the naked singularity is absent.
However, $C_+$ depends on the matter state (see (4.15)), and for narrow matter
pulses $C_+$ is negative, resulting in the appearance of the naked 
singularity and  the pathological behavior of the Hawking flux. The authors
of \cite{vw} have conjectured that the back-reaction prevents naked 
singularities to form. The $C_+ <0$ solution seem to be a counterexample
to this conjecture. However, one has neglected the higher order corrections,
so it is still possible that in the full non-perturbative solution the naked
singularities do not appear in the semi-classical regions. 
Also it is very indicative from (4.25) that $C_+$
is positive for reasonable matter pulses, i.e. pulses which are wider than the
short-distance cutoff $l_c$, which can play the role of a 2d Planck length.

\sect{Acknowledgments}

\noindent We would like to thank S. Bose, H. Kastrup, T. Strobl and 
A. Bogojevi\'c for useful discussions. A.M. would also like to thank Serbian
Ministry of Science and Technology, Swedish Institute and  Chalmers
Institute of Theoretical Physics for financial support.

\end{document}